\documentclass[11pt]{article}
\usepackage{fleqn,cospar}
\usepackage{url}
\usepackage{graphicx}
\usepackage{natbib}
\usepackage{amssymb}

\title{CYGNUS X-1 FROM RXTE: MONITORING THE SHORT TERM VARIABILITY}

\author{K.~Pottschmidt\address{Institut f\"ur Astronomie und Astrophysik,
    Waldh\"auser Str. 64, D-72076 T\"ubingen, Germany},
  J.~Wilms$^{1}$, R.~Staubert$^{1}$,
  M.A.~Nowak\address{JILA, University of Colorado, Boulder, CO
    80309-440, USA}, 
  W.A.~Heindl\address{CASS, University of California San Diego, La
    Jolla, CA 92093, USA}, and
  D.M.~Smith\address{SSL, University of California Berkeley, Berkeley,
    CA 94720, USA}} 

\begin{document}

\maketitle

\begin{abstract}
  We present temporal and spectral results from monitoring
  \mbox{Cygnus~X-1} with the Rossi X-ray Timing Explorer (RXTE) in 1998 and
  1999. We concentrate on the long term evolution of the hard state timing
  properties, comparing it to the 1996 soft state evolution. This leads to
  the following results: 1. the hard and soft state time lag spectra are
  very similar, 2. during state transitions, the lags in the 1--10\,Hz
  range increase by more than an order of magnitude, 3. in the hard state
  itself, flaring events can be seen --- the temporal and spectral
  evolution during the flare of 1998 July identifies it as a ``failed state
  transition''.  During (failed) state transitions, the time lag spectra
  and the power spectra change predominantly in the 1--10\,Hz range. We
  suggest that this additional variability is produced in ejected coronal
  material disrupting the synchrotron radiation emitting outflows present
  in the hard state.
\end{abstract}

\section*{INTRODUCTION AND DATA ANALYSIS}

Galactic black hole candidates (BHC) are predominantly found in two states:
the hard state, in which the X-ray spectrum is a Comptonization spectrum
emerging from a hot electron cloud with a typical electron temperature of
$\sim$150\,keV \citep{dove:97b,poutanen:98a}, and the soft state, in which
the X-ray spectrum is thermal with a characteristic temperature of $kT_{\rm
  BB}\lesssim 1$\,keV to which a steep power-law is added \citep[and
references therein]{cui:96a,gierlinski:99a}. The X-ray states are also
known to be associated with radio states \citep[see, e.g.,][for a
review]{fender:00b}.  The canonical BHC \mbox{Cyg~X-1} stays predominantly
in the hard state, and only occasionally transits into the soft state for a
few months \citep[][see also
Fig.~\ref{fig:softlag}]{gierlinski:99a,cui:96a,cui:97c}.

In addition to spectral analysis, timing analysis can also yield insight
into the radiation mechanisms in BHCs. Apart from power spectrum
analysis \citep{belloni:90a,gilfanov:99b,churazov:00a}, higher order
statistics like the frequency-dependent coherence function and time lags
have proven to be useful in evaluating accretion models
\citep{hua:98a,nowak:98a}. Although very different in their specific
physical assumptions, most models for the hard and the soft state assume
that the observed time lags are an indicator of some characteristic
geometrical size of the Compton corona. In the case of \mbox{Cyg~X-1}, the
apparent decrease of the X-ray lags during the 1996 soft state transition
of the source was considered evidence that the size of the Comptonizing
corona during the soft state is smaller than during the hard state
\citep{cui:97c}. First comparisons of transition and soft state lags to
hard state lags, however, indicated that the physical interpretation has to
be more complex \citep{cui:99a}.

In 1998 we initiated a monitoring campaign of \mbox{Cyg~X-1} with RXTE to
systematically study long term variations of the hard state properties over
a period of several years. Previous results have been presented elsewhere
\citep{pottschmidt:00b}. In 1998 weekly pointings of $\sim$3\,ks duration
and in 1999 two-weekly pointings of $\sim$10\,ks duration were performed.
In this contribution we present results of the analysis of these
observations.  Together with archival observations from 1996, we consider
103 observations.

We concentrate on the data from the Proportional Counter Array
\citep[PCA;][]{jahoda:96b}, and use the RXTE All Sky Monitor
\citep[ASM;][]{remillard:97a} to place our observations in the context of
Cyg~X-1's long term evolution.  The data were reduced using the procedures
described in detail elsewhere \citep{wilms:98c}. We mainly show results
obtained from lightcurves with a resolution of 4 or 16\,ms observed in two
energy bands --- $\leq$4\,keV and $\sim$8--13\,keV. The chosen energy bands
are known to be well suited for time lag studies of this source
\citep{nowak:98a}. The exact bands used to compare the 1996
data to the monitoring data were the closest matches possible for the
available PCA modes. Furthermore, we describe the results of the spectral
analysis of standard PCA spectra in the 2--20\,keV range.  We present the
long term evolution of the averaged time lags, the frequency dependent time
lag spectra and power spectra, and the spectral evolution during the flare
of 1998~July.  Finally, we discuss these results in the light of current
accretion models.

\section*{LONG TERM EVOLUTION OF THE TIME LAGS}

\begin{figure}
\centerline{\includegraphics[width=0.75\textwidth]{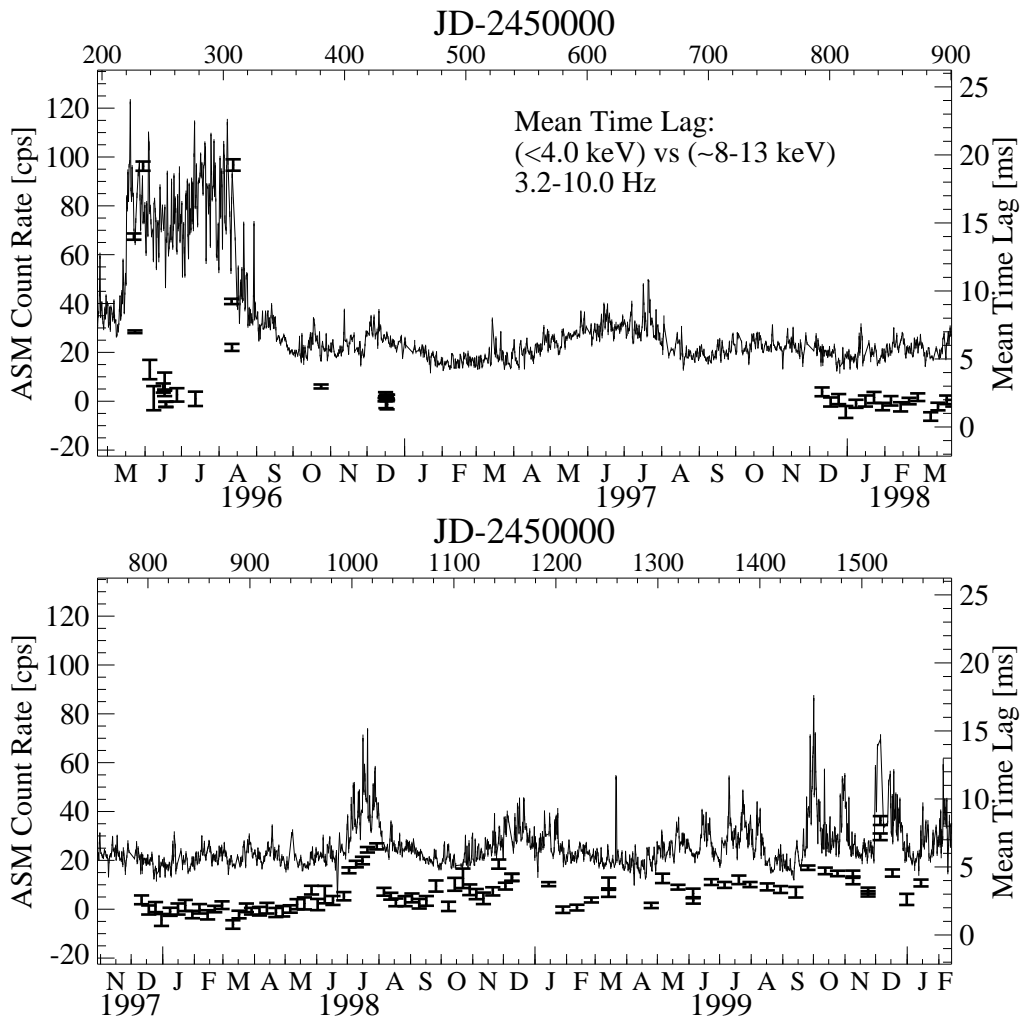}}
\vspace{-0.8cm}
\caption{\label{fig:softlag}RXTE ASM count rate (line) and mean time lag
  (symbols) between the energy band below 4\,keV and the energy band from 8
  to 13\,keV in the frequency range from 3.2 to 10\,Hz for the 1998/1999
  monitoring campaign as well as for the RXTE data obtained during the 1996
  soft state. During hard state phases there is a high correlation between
  the soft X-ray luminosity and the time lag.}
\end{figure}

\begin{figure}[t]
\centerline{\includegraphics[width=0.7\textwidth]{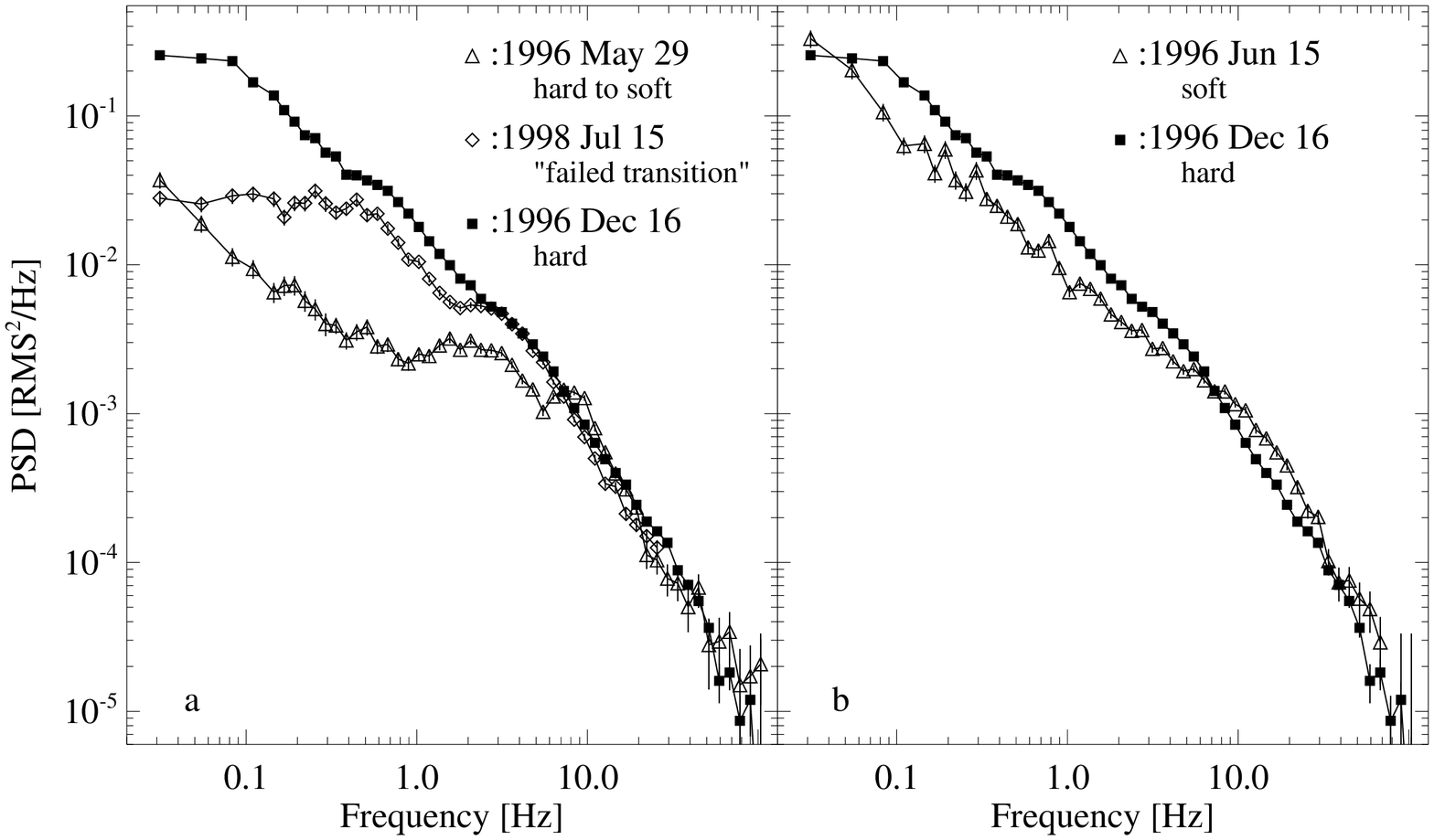}}
\centerline{\includegraphics[width=0.7\textwidth]{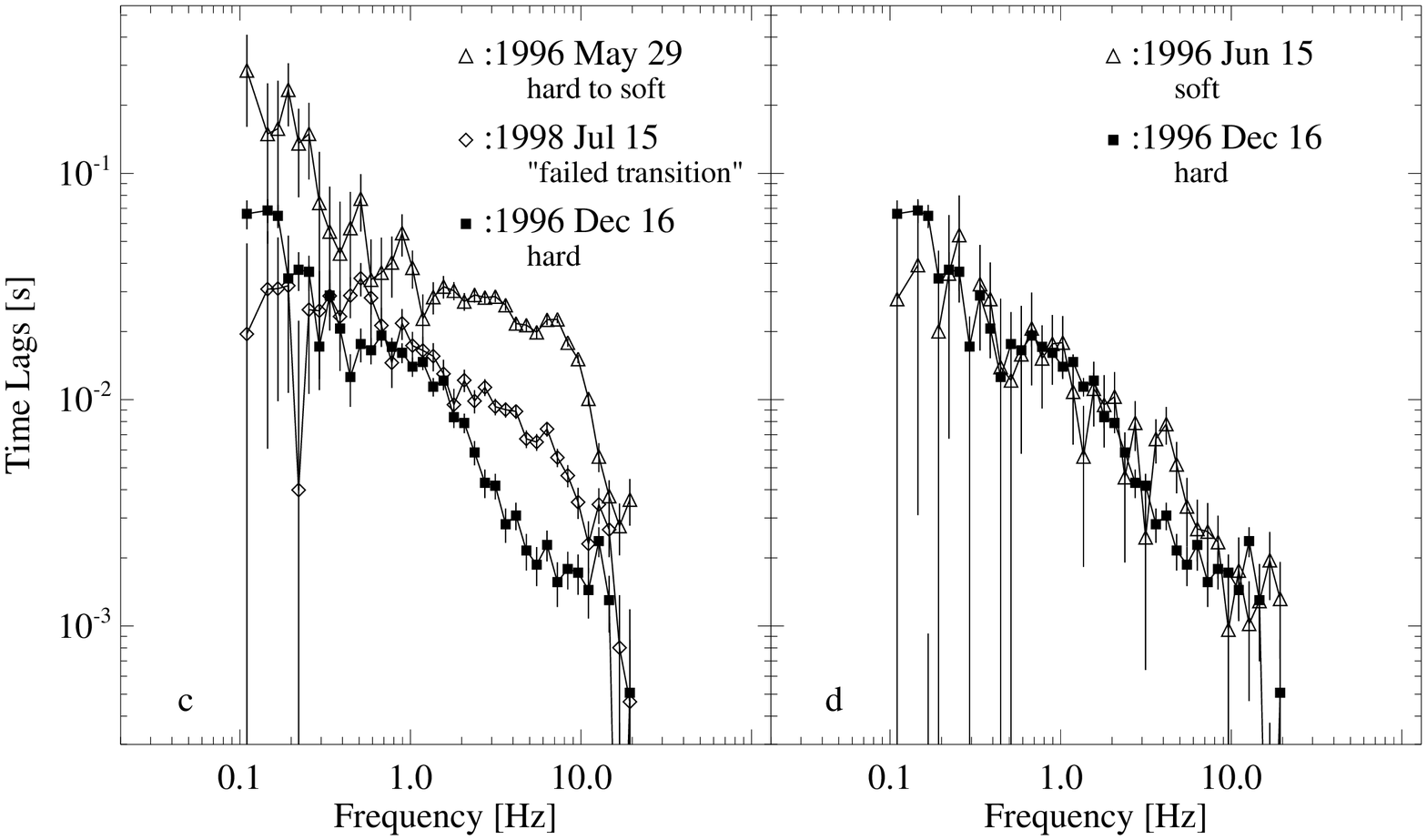}}
\vspace{-0.8cm}
\caption{\label{fig:flare}Comparison between the power spectral densities
  (8--13\,keV) and time lags for  
  the 1996 state transition, a characteristic 1996 hard state observation,
  and the 1998~July failed state transition (subfigures a and~c). Subfigures b
  and~d compare the PSDs and lags between the hard and the soft state. }
\end{figure}
In Fig.~\ref{fig:softlag}, we display the daily averaged RXTE ASM count
rate superposed on the mean time lag between the soft and the hard energy
bands in the frequency band from 3.2 to 10\,Hz. We have found that the
X-ray time lag is most variable in this frequency band
\citep{pottschmidt:00b}. In the lower panel, the 1998/1999 monitoring is
displayed. It shows that in the hard state there is a clear correlation
between the soft X-ray luminosity and the mean time lag (Pearson's
correlation coefficient of 0.78). This is especially true during ``flaring''
events (e.g., in 1998 July, see below).

In the light of current accretion models, however, we would expect the lag
to decrease as the source softens (with the corona becoming smaller). This
was thought to happen during the 1996 full state transition, where the lag
was seen to be considerably smaller in the soft state than immediately
before and after \citep{cui:97c}. Considering the hard state monitoring
observations as well, however, we see a different picture (upper panel of
Fig.~\ref{fig:softlag}): While the absolute values of the lags are
comparable in the hard state and in the soft state, they are significantly
longer during the transitions. In fact, the frequency dependence of the lag
is very similar for the soft and hard states (Fig.~\ref{fig:flare}d). It is
interesting to note that for the hard energy band (8--13\,keV) and the
frequency range considered here, the timing behavior as characterized by
the power spectral density (PSD) also looks similar
(Fig.~\ref{fig:flare}b). A qualitative model for explaining this PSD
phenomenology in terms of transfer of variability structures through the
disk/corona system has recently been presented by \citet{churazov:00a}.

Previous analyses of two hard state observations already suggested that the
soft and hard state lags might not be so different as previously thought
\citep{cui:99a}. Our numerous hard state monitoring observations now
clearly indicate that \emph{the X-ray lag spectrum of \mbox{Cyg~X-1} is
  rather independent of the spectral state}. The \emph{enhanced lags are
  associated with transition or flaring intervals}, and not with the state
of the source itself.

\section*{FLARING \& HARD STATE STABILITY}

Of special interest are the episodes of enhanced ASM count rates in the
hard state, which, as we have seen, also show larger mean time lags than
the ``canonical'' hard state. Figs.~\ref{fig:flare}a and~c display the PSD
and the time lag spectrum for an observation performed during the 1998 July
flare and compares them to a typical hard state observation as well as to
the transition phase of the 1996 state transition.

Compared to the typical hard state lag spectrum the appearance of an
additional component in the 1 to 10\,Hz range can be seen during the
1998~July flare. This component becomes more prominent and ``flattens''
during the hard to soft transition in 1996~May, with the lag being longer
by almost a factor of 10 at 6\,Hz as compared to 1996 December.  As has
been reported by \citet{cui:97c}, the PSDs of the state transition show an
additional steep component at lower frequencies, a shift of the break
frequency to higher frequencies, and additional ``QPO-like'' features.  The
PSD of the 1998 flare is intermediate between the hard state and the state
transition, i.e., the break frequency has moved from $\sim0.1$\,Hz to
$\sim0.5$\,Hz. Furthermore, a quasi-periodic oscillation appears at
$\sim2$\,Hz, i.e., in the frequency range that exhibits the enhanced lag
(the QPO might also be present in the hard state PSD at $\sim0.4$\,Hz).
Since the flaring episode displays all the features of the real state
transition, presumably at the onset of their appearance, we identify it as
a ``failed state transition''. 

This interpretation as a failed state transition is confirmed by the
spectral analysis: Outside of the ``flaring episode'' the hard state X-ray
spectrum as seen by the PCA can be described by the canonical power-law
(photon index $\Gamma=1.80\pm0.01$)\footnote{All uncertainties are at the
  90\% level.}, reflected off neutral material with a covering angle of
$\Omega/2\pi= 0.44\pm0.04$ (the inclination has been fixed at $35^\circ$).
An additional broad ($\sigma=0.8$\,keV) Fe line at 6.4\,keV with an
equivalent width of 150\,eV is required ($\chi^2/{\rm dof}=19.6/42$; note
that the $\sim 1$\% systematics of \citealt{wilms:98c} have been applied).
During the flare the spectrum softens to $\Gamma=2.13\pm 0.02$ and the
covering factor increases to $\Omega/2\pi=0.93\pm0.07$. At low energies an
additional soft spectral component appears (modeling this component as a
multi-temperature disk black body results in $kT_{\rm
  in}=0.45\pm0.05$\,keV. The soft component contributes $\sim$7\% of the
flux at 3\,keV. Thus, although the power law component still dominates the
higher energy spectrum, there is clear evidence for a strong soft
component, such that the 1998~July event should be associated with a failed
state transition.

\begin{figure}
\centerline{%
\includegraphics[width=0.38\textwidth]{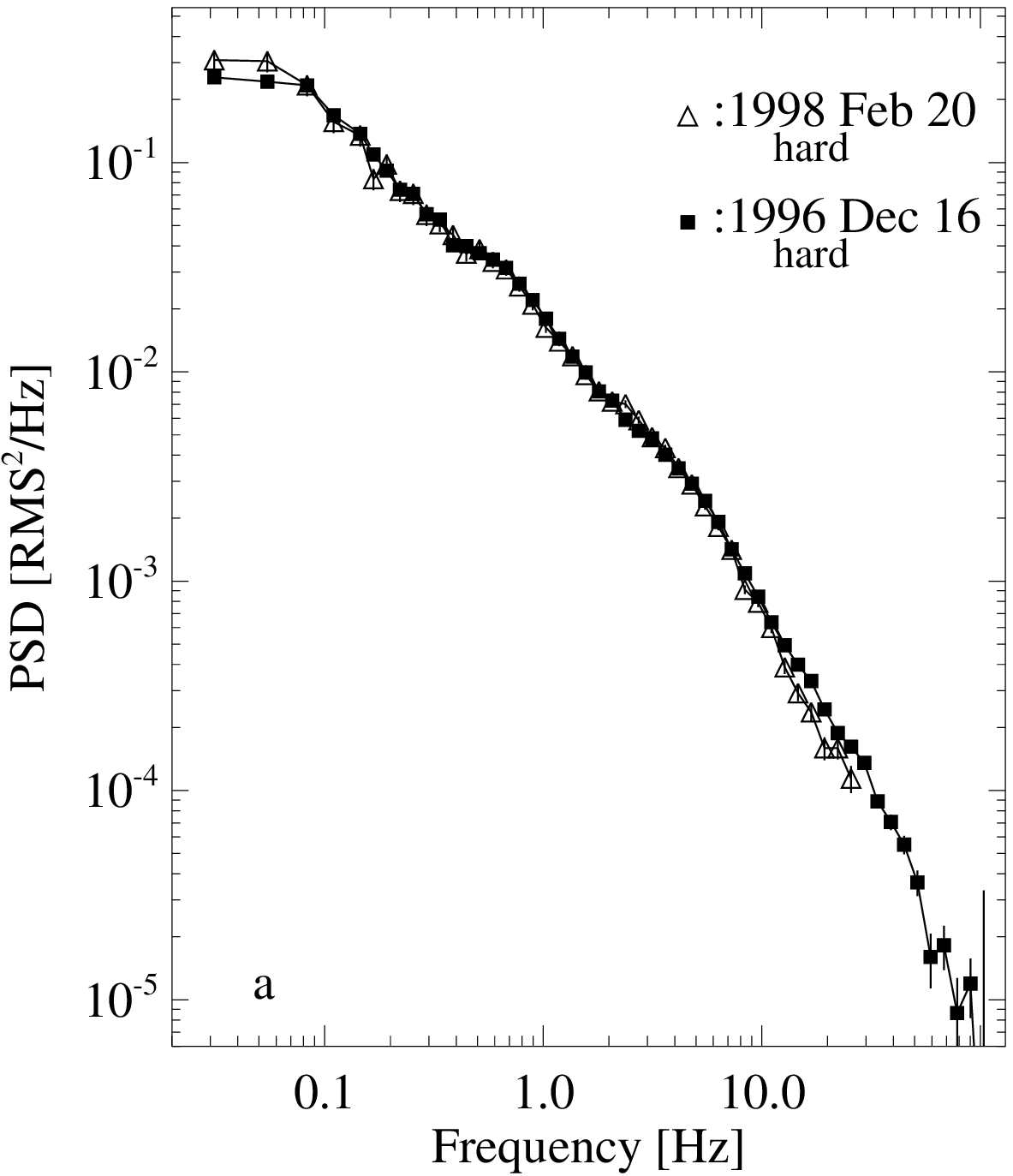}
\hspace*{2mm}
\includegraphics[width=0.38\textwidth]{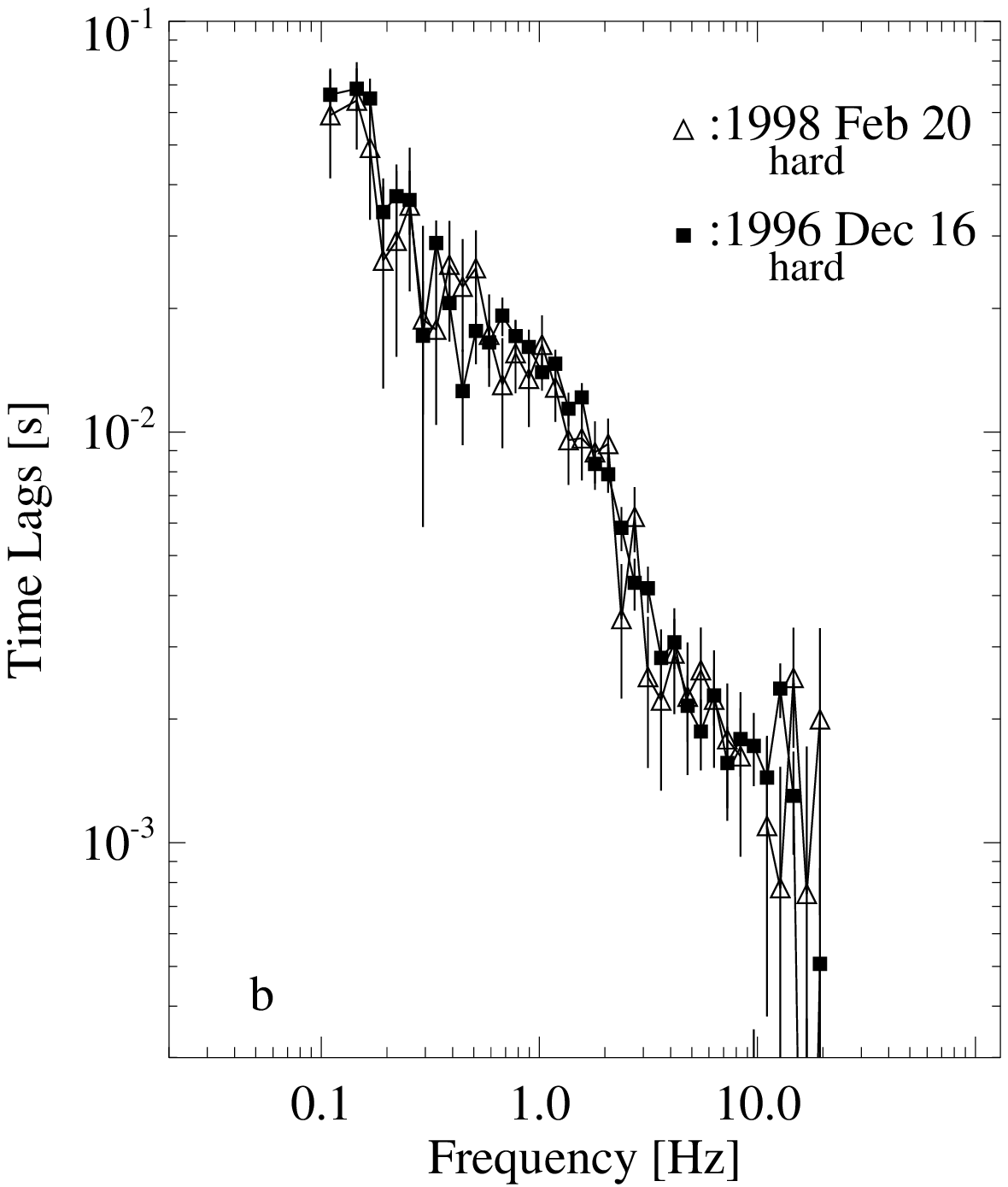}
}
\vspace{-0.8cm}
\caption{\label{fig:hard}Comparison of the PSDs and time lags for two hard
  state observations separated by 14~months. Both indicators for the
  temporal variability are virtually indistinguishable.}
\end{figure}

Although the flaring episodes introduce considerable long term variability
into the hard state properties, the timing characteristics can be
reproduced to great detail in observations separated by more than a year
(Fig.~\ref{fig:hard}): the time lag spectra and the PSDs of the
1996~December and the 1998~February observation are identical.  This
includes the ``shelves'' in the lags and the ``wiggles'' in the PSDs that
might indicate that multiple processes are responsible for the observed
timing behavior \citep[see also][]{nowak:00b}.

\section{DISCUSSION}
Since Comptonization models generally give a satisfactory explanation for
the X-ray spectrum, most models for the generation of the X-ray time lags
assume that the lags are at least partially produced by scattering of seed
photons (which might have an intrinsic time lag) in a Compton corona
\citep[and references therein]{poutanen:98a}. Comptonization cannot be the
only cause for the time lags, however, as the magnitude of the lag implies
very large Compton coronae \citep[$\gtrsim$300 gravitational
radii][]{nowak:98a}, in conflict with the assumption of most Comptonization
models.  Furthermore, the energy dependence of the lags is also in
disagreement with Comptonization \citep{nowak:98a}.
 
We note, however, that the coherence function in the frequency regime
considered here is not equal to unity, but slightly smaller. This might be
an indication that (part of) the time lag is produced by non-linear
processes in the accretion disk, and not by the time delay due to the
Compton scatterings in the corona (Churazov, 2000, priv.\ comm.).

Nevertheless, although all of these arguments point against Comptonization
as the sole cause for the X-ray lags, one is tempted to assume that the
magnitude of the lag is at least somehow proportional to the characteristic
size of the X-ray emitting region. Accepting this paradigm, however,
implies that the size of the X-ray emitting corona during the soft and hard
states is the same as the lags during these states are identical. This is
puzzling, given that the luminosity of the corona, and thus the release
mechanism for the potential energy of the accreted material, is so
different between the hard and the soft state.

During state transitions, larger X-ray lags are observed in the 1--10\,Hz
band. These larger lags might be related to the radio outbursts seen during
the transitions \citep{pottschmidt:00b}. Galactic BHCs in the hard state
emit optically thick radio emission, while the soft state is radio quiet.
During transitions, the radio emission tends to be optically thin and
highly variable (e.g., GX~339$-$4, \citealt{corbel:00a}, but also Cyg X-1
itself; see \citealt{fender:00b} for a review). For Cyg~X-1, there exists a
good correlation between the radio and the soft X-ray flux
\citep{brocksopp:99b}, and state transitions are correlated to radio
outbursts \citep[see also \citealt{tananbaum:72b}]{zhang:97e}.  Such
outburst are usually associated with the ejection of synchrotron radiation
emitting outflows \citep[see][for the case of
GX~339$-$4]{fender:99b,corbel:00a}. We suggest that the radio outflows and
the larger lags might be related \citep{pottschmidt:00b}. One possible
source for the electrons seen in the radio outflow might be the accretion
disk corona. During state transitions, when the accretion disk is thought
to reconfigure, the corona might be ejected and then observed as the radio
outflow. In this picture, soft X-rays from the accretion disk might get
Compton upscattered in the base of this outflow. As the base is much larger
than the common accretion disk corona, one would expect larger X-ray lags.
We note that such a picture, where part of the observed lags is due to the
presence of a jet, has been previously suggested by van Paradijs (1999,
priv.\ comm., see also \citealt{fender:99b}).

\section*{ACKNOWLEDGEMENTS}
This work has been supported by DFG grant Sta 173/22, by DLR grant 50
OO 9605, by NASA grants NAG5-3072, NAG5-3225, and NAG5-7265, and by travel
grants from the DAAD and the NSF.

\end{document}